# Experimentation for Packet Loss on MSP430 and nRF24L01 Based Wireless Sensor Network

**S. S. Sonavane, Dr. B. P. Patil**
Department of Electronics Engineering,
Maharashtra Academy of Engineering, Alandi (Pune), State-Maharashtra (India)
Email: sssonavane@gmail.com, bpatil@maepune.com
**Dr. V. Kumar**
Department of Electronics & Instrumentation Engineering,
Indian School of Mines University, Dhanbad-826004, State-Jharkhand (India)
Email: vkumar52@hotmail.com

**-------------------------------------------ABSTRACT---------------------------------------**
In this paper, a new design of wireless sensor network (WSN) node is discussed which is based on components with ultra low power. We have developed a Low cost and low power WSN Node using MSP430 and nRF24L01. The architectural circuit details are presented. This architecture fulfils the requirements like low cost, low power, compact size and self-organization. Various tests are carried out to test the performance of the nRF24L01 module. The packet loss, free Space loss (FSL) and battery lifetime calculations are described. These test results will help the researchers to build new applications using above node and to work efficiently with nRF24L01.

**Keywords:** Wireless Sensor Networks, MSP430, nRF24L01, Free Space Loss, battery lifetime, packet loss.

## 1. INTRODUCTION

A wireless sensor network is a network made up of hundreds or thousands of Sensor nodes which are densely deployed in an unattended environment [1]. These nodes are capable of communicating by means of wireless communications, sensing and self-computation (software, hardware, algorithms). Hence the wireless sensor network is the result of the combination of sensor, embedded techniques, distributed information processing, and communication mechanisms. The sensor network is more application specific than traditional networks designed to accommodate various applications [2,3,4]. The organization and architecture of a sensor network should be designed or adapted to suit a special task so as to optimize the system performance, maximize the operation lifetime and minimize the cost.

The WSN node consists of three components i.e. microcontroller, RF module, sensor and battery [5]. As the nodes are placed in remote places the battery lifetime plays a vital role. Hence to reduce the power consumption, we use ultra low power consumption based components. Also the duty cycle is kept < 1% so that the on time is reduced [6]. To test the performance of radio, we had carried out tests for packet loss, free Space loss (FSL) and battery lifetime.

The paper is organized as: section II describes the design of WSN node. The section III describes the test setup and procedure for packet loss, free Space loss (FSL) and battery lifetime calculations and the corresponding graphs are explained. Section IV concludes the paper.

## 2. DESIGN OF ULTRA LOW POWER WSN NODE

The Fig.1 shows WSN node with following components:

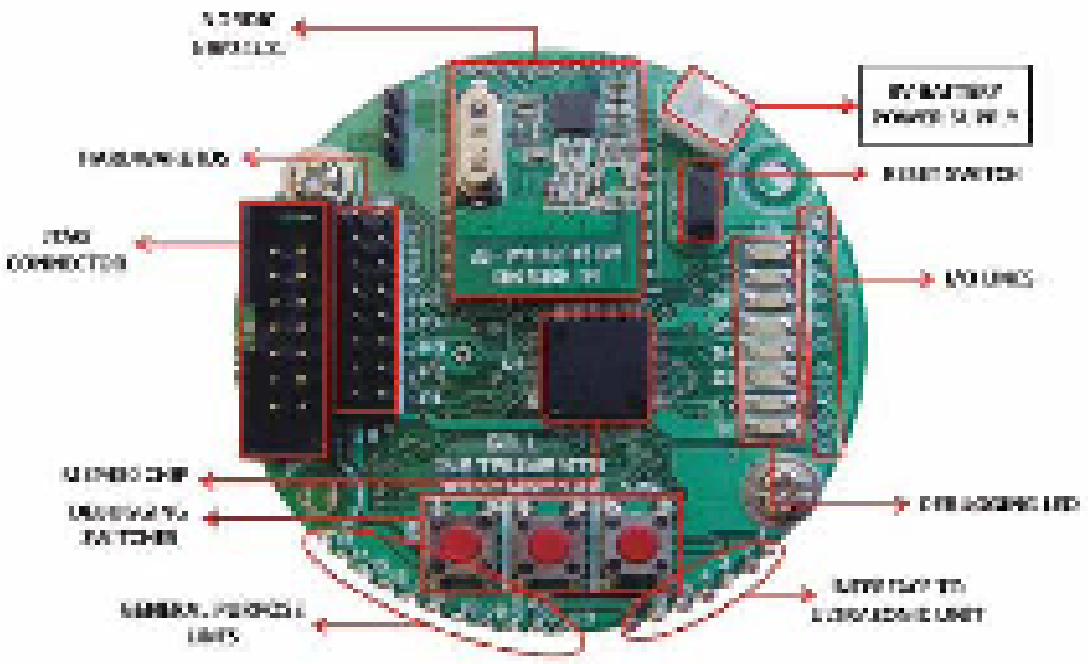

Fig.1 Front side of WSN Node with MSP430 and nRF24L01

### 2.1 MSP430F1612

This Node features MSP430F1612, ultra low power microcontroller having the lowest power consumptions and fastest wake-up cycles. The Texas Instruments MSP430 family of ultra low-power microcontrollers consists of several devices featuring different sets of peripherals targeted for various applications. The architecture, combined with five low power modes is optimized to achieve extended battery life in portable measurement applications. The device features a powerful 16-bit RISC CPU, 16-bit registers, and constant generators that attribute to maximum code efficiency [7-8].

The digitally controlled oscillator (DCO) allows wake-up from low-power modes to active mode in less than 6µs and may operate up to 8MHz. Typically, the DCO will turn on from sleep mode in 300ns at room temperature. The MSP430F1612 has two built-in 16-bit timers, a fast 12-bit A/D converter, dual 12-bit D/A converters, one or two universal serial synchronous/ asynchronous communication interfaces (USART), I2C, DMA, and 48 I/O pins. The core module also has a 4 Mbit flash chip that can be used for storing several firmware images or for logging data.

### 2.2 The nRF24L01 RF Module

The nRF24L01 is a single chip radio transceiver for the global, license-free 2.4 GHz ISM band [9]. The low cost nRF24L01 is designed to merge very high speed communications (up to 2Mbit/s) with extremely low power (the RX current is just 12.5mA). The transceiver consists of a fully integrated frequency synthesizer, a power amplifier, crystal oscillator, demodulator, modulator and Enhanced ShockBurst protocol engine. In addition, the nRF24L01 also offers an innovative on-chip hardware solution called 'MultiCeiver' that can support up to six simultaneously communicating wireless devices. This makes it ideal for building wireless Personal Area Networks in a wide range of applications.

Output power, frequency channels, and protocol set-up are easily programmable through an SPI-bus. Current consumption is very low, only 8.5mA at an output power of -6dBm and 12.5mA in RX mode. Built-in modes such as Power Down (400nA current) and Standby (32µA at 130µs wakeup), makes significant power savings easily realizable. The data rate can be chosen between 1 and 2Mbit/s. This allows for short time-on-air and therefore low power consumption.

### 2.3 Nordic and MSP430 Connections

The nRF24L01 is a single chip 2.4 GHz Transceiver. Nordic transceiver has total nine external connectors in which six are connected to the MSP430. The nRF24L01 is interfaced with MSP430 using SPI (Serial Peripheral interface) at port3 and synchronization between two has been achieved by MCLK (Master clock).

SPI specifies four signals: clock (SCLK), master data output slave data input (MOSI), master data input slave data output (MISO) and slave select (CSS). SCLK is generated by the master and input to all slaves. MOSI carries data from master to slave. MISO carries data from slave back to master. A slave device is selected when the master asserts its CSS signal.

### 2.4 Transmitted Packet format

The nRF24L01 is configured and operated through a Serial Peripheral Interface (SPI.) Through this interface the register map is available. The register map contains all configuration registers in the nRF24L01 and is accessible in all operation modes of the chip. The transmitted packet is shown in Fig. 2.

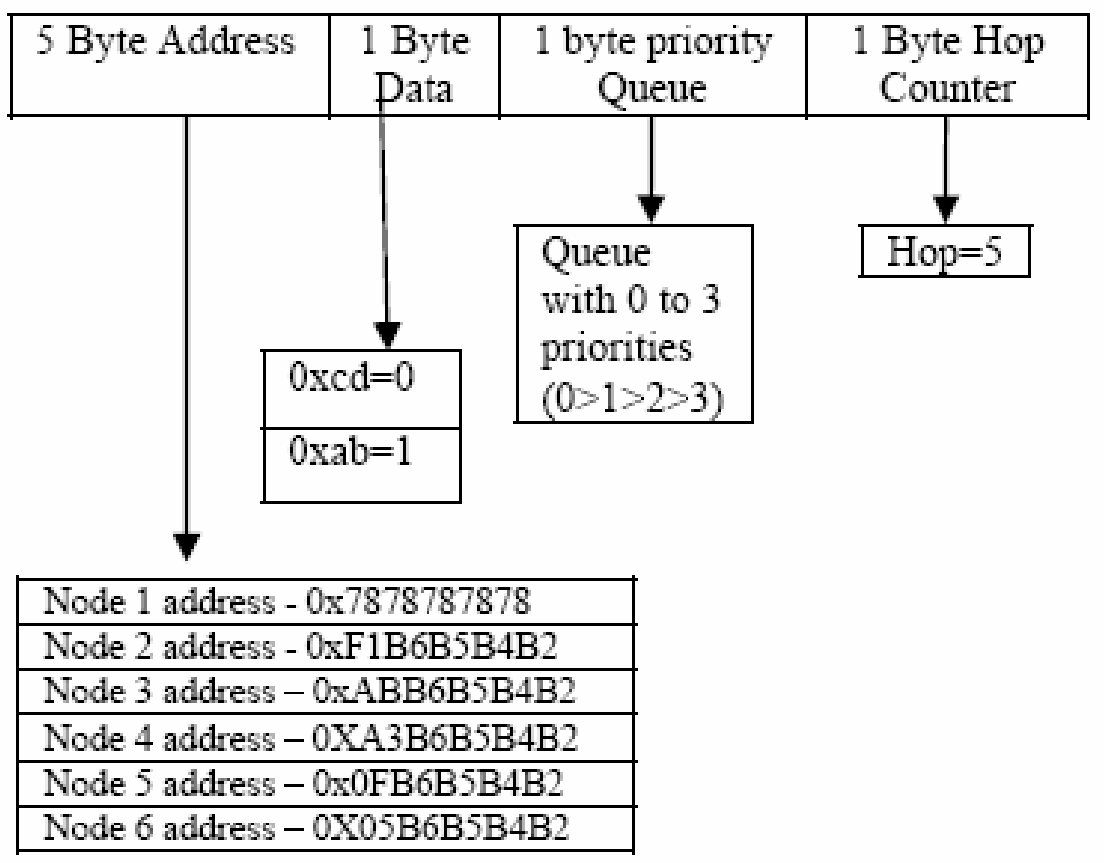

Fig.2 Transmitted Packet Format of nRF24L01

The embedded baseband protocol engine (Enhanced ShockBurst) is based on packet communication and supports various modes from manual operation to advanced autonomous protocol operation. Internal FIFOs (First In First out) ensure a smooth data flow between the radio front end and the system's MCU. Enhanced ShockBurst reduces system cost by handling all the high-speed link layer operations.

The radio front end uses GFSK (Gaussian frequency Shift Keying) modulation. It has user configurable parameters like frequency channel, output power and air data rate. The air data rate supported by the nRF24L01 is configurable to 2Mbps. The high air data rate is combined with two power saving modes which makes the nRF24L01 very suitable for ultra low power designs. Internal voltage regulators ensure a high Power Supply Rejection Ratio and a wide power supply range. In power down mode nRF24L01 is disabled with minimal current consumption.

In power down mode all the register values available from the SPI are maintained and the SPI can be activated. Power down mode is entered by setting the PWR_UP bit in the CONFIG register low. By setting the PWR_UP bit in the CONFIG register to 1, the device enters standby-I mode. Standby-I mode is used to minimize average current consumption while maintaining short start up times. In this mode part of the crystal oscillator is active. This is the mode the nRF24L01 returns to from TX or RX mode when CE is set low.

Enhanced ShockBurst is a packet based data link layer. It features automatic packet assembly and timing, automatic acknowledgement and retransmissions of packets. Enhanced ShockBurst enables the implementation of ultra low power, high performance communication with low cost host microcontrollers. The features enable significant improvements of power efficiency for bi-directional and unidirectional systems, without adding complexity on the host controller side. The main features of Enhanced ShockBurst are:

- 1 to 32 bytes dynamic payload length
- Automatic packet handling
- Auto packet transaction handling
- Auto Acknowledgement
- Auto retransmission
- 6 data pipe MultiCeiver for 1:6 star networks

## 3. TESTING DETAILS

### 3.1 To measure Packet Loss:

The test setup for a packet loss measurement is shown in Fig 3. We transmit desired no. of packets from Node1. At node 2 we count the received Packets to determine the Packet Error Rate (PER). The procedure is as follows:

1. Transmit RF signal from Node1 that the receiver can demodulate.
2. Vary the No. of Packets in a controlled way at Node1.
3. Count the number of errors in the received Packets at Node 2.
4. Repeat above step for different distances and different packets.

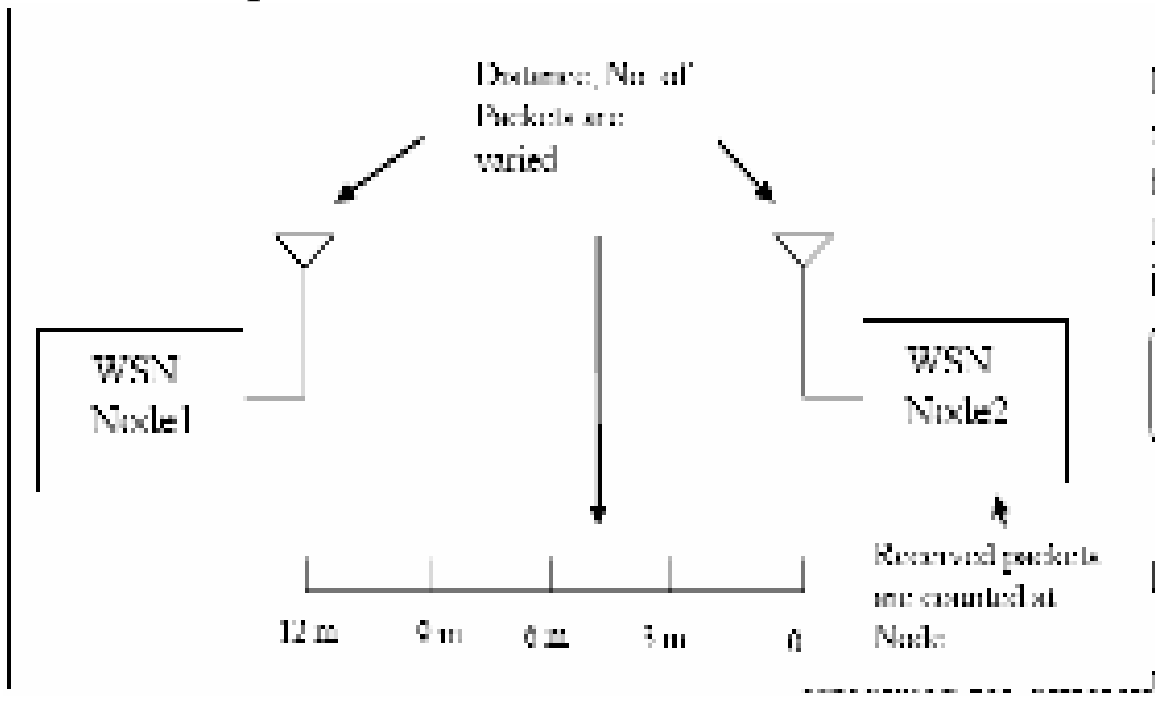

Fig. 3 Test setup for Packet Loss measurement

When conducting Packet loss measurements the receiver is tested on its performance limit. Low signal levels make this test the most demanding of all RF prototype tests. It is very important to control the test environment closely as any additional noise directly affects the measurement. So place the WSN Node under test into the area with good line of sight.

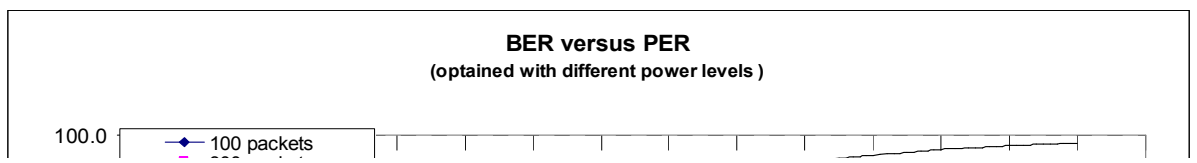

Fig.4 Graph of BER Vs PER for different packet size

For Packet loss measurements we program the node to transmit the packets from 100 packets to 5000 packets shown in Fig 4. The packets are formed with the format as explained in Fig.2. These packets are having known payload. Hence at receiver these payloads are compared with the known data and the packet bit error rate can be calculated. The graph of BER Vs PER is shown Fig.5.

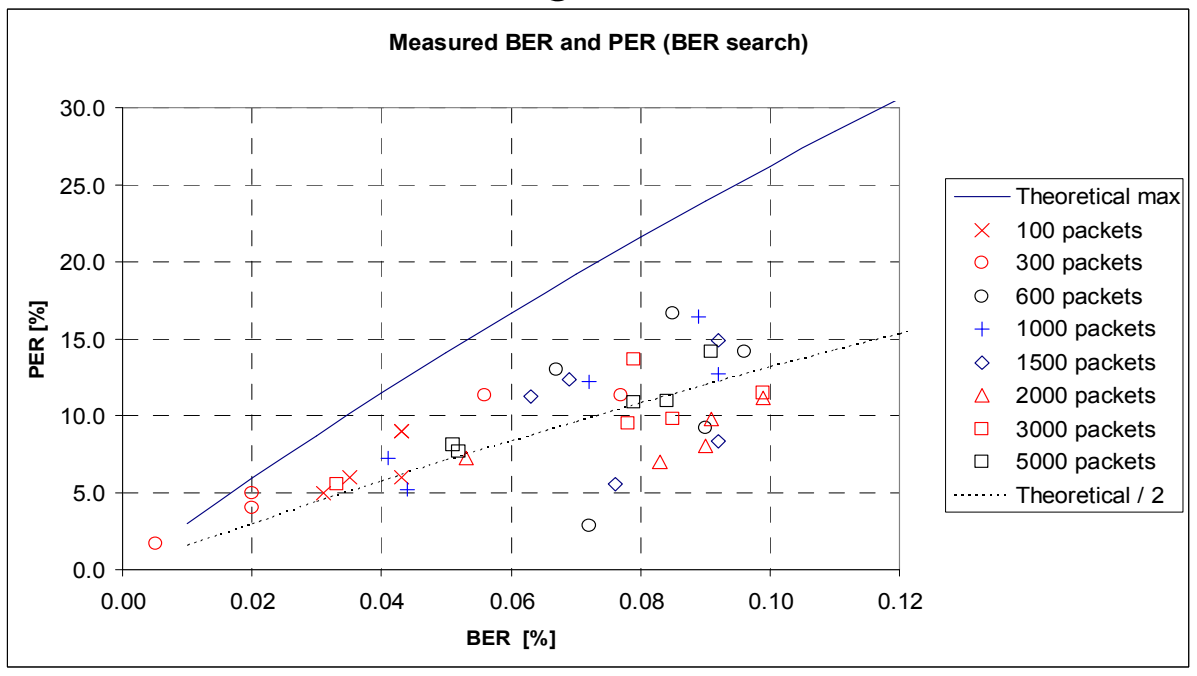

Fig.5 Graph of BER Vs PER

### 3.2 Calculation of FSL from Range:-

With ¼ wave PCB whip antenna gain of -5 to -10 dB you can achieve an effective radiated power (ERP) from the transmitters of < –5dBm @ 0 dBm Transmitter (TX) output power and a receiver sensitivity of > –75 dBm. In other words the free space loss (FSL) in the system is < 70 dB.

Free space loss (FSL) is given by:

$$FSL = -20 \text{ Log} \left( \frac{\lambda}{4 \pi R} \right)$$

Where λ is the wavelength at 2.4 GHz and R is the maximum theoretical line of sight range obtainable in meters. The FSL is calculated for different range according to the above formula as given in Table 1. The Fig 6 shows graph of Range Vs FSL in which the FSL increases with distance logarithmically.

Table 1 Range and FSL details

| Range (meter) | 1 | 2 | 3 | 4 | 5 | 6 | 7 | 8 | 9 | 10 |
|---|---|---|---|---|---|---|---|---|---|---|
| FSL (dB) | 40 | 46 | 49.6 | 52.08 | 54.02 | 55.61 | 56.95 | 58.1 | 59.13 | 60.04 |

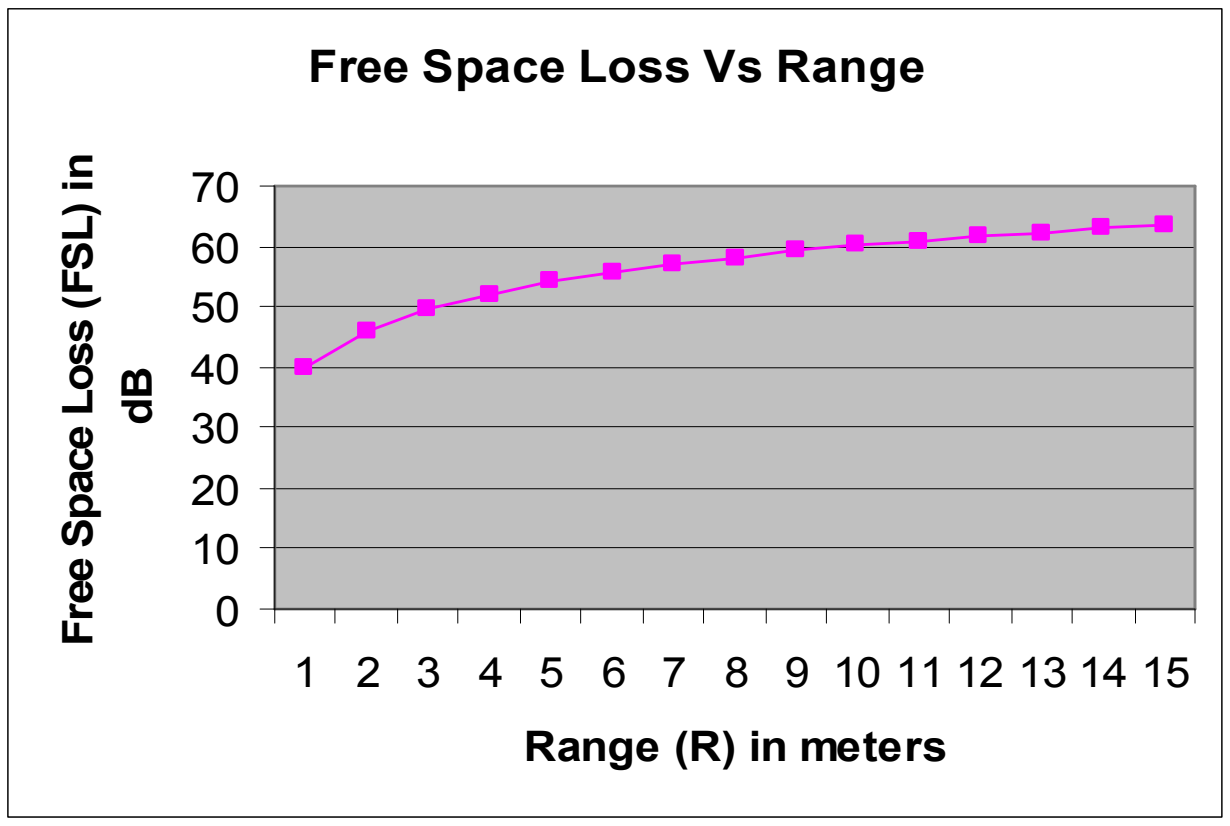

Fig. 6 Graph of range Vs FSL

### 3.3 Battery Life calculations

Choosing the right battery can determine the success or failure of a wireless sensor networking project. We select the Primary lithium-thionyl chloride (Li-$SOCl_2$) LST 14500 for our node [10]. The benefits of this battery are High voltage response, stable during most of the lifetime of the application, easy integration in compact system, Non-flammable electrolyte, Low self-discharge rate (less than 1% after 1 year of storage at + 20°C) and Underwriters Laboratories (UL) Component Recognition (File Number MH 12802) . The rating is 2450 mAh.

So the battery life can be calculated by taking into account the parameter like the current consumptions of MSP430 and nRf24L01. Table 2 shows the details of battery lifetime. The duty cycle also decides the wakeup and sleep modes of MSP430 and nRF24L01. The Nordic has settings for four different power levels through PWR_UP register. The Fig. 7 shows that at low power levels the battery life will be more.

Table 2 Calculation of Battery lifetime for our WSN Node

| Parameter | Settings | Unit |
|---|---|---|
| Overhead | 65 | bits |
| Payload length | 8 | bits |
| Packet length | 73 | bits |
| Bit rate | 2000000 | bits/s |
| Time on air | 0.3 | s |
| Time in RX | 0.00003252 | s |
| MCU+TX Current | 11.6 | mA |
| MCU+RX current | 12.9 | mA |
| PLL- Lock time | 0.00013 | s |
| PLL- Lock TX current | 8 | mA |
| PLL- Lock RX current | 8.4 | mA |
| Power_Dn current | 0.0009 | mA |
| Duty-cycle period | 55.7 | s |
| Power_up current | 0.285 | mA |
| Power_up time | 0.0015 | s |
| I(avg) | 0.06342619 | mA |
| | | |
| Battery Used | 2450 | mAh |
| | | |
| Lifetime | 38627.5768 | h |
| | 1609.48237 | days |
| | **4.42165486** | **years** |

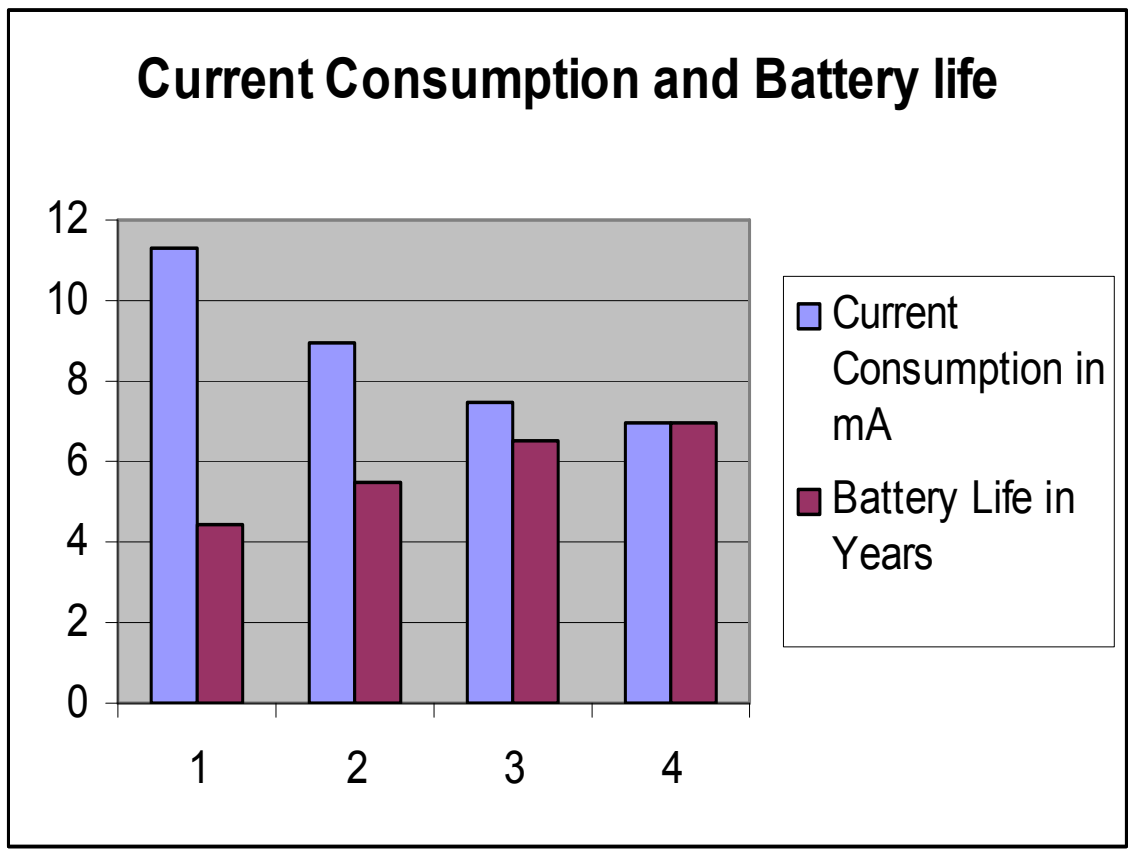

Fig. 7 Graph of Current Consumption and battery Lifetime

## 4. CONCLUSION

The hardware designed is a low power solution for WSN as compared to other commercially available motes. The test result shows that the design is having the packet loss about 0.1 %. The packet loss increases with the transmission range. The battery lifetime is also dependent on the power setting at transmitter. Hence in applications if the nodes are close to each other, use low power level (i.e. -18 dBm). This will enhance the battery lifetime.

This ultra low power design can be effectively adopted for applications such as environmental monitoring, home automation, Car park management system etc.

## Authors Biography

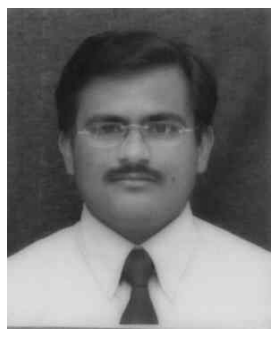

**Sonavane S. S.** is currently pursuing his Ph.D. from Indian School of mines, Dhanbad. He is currently working as Professor in Dept. of Electronics Engineering at Maharashtra Academy of Engineering, Alandi, Pune. He has published around 26 papers in International and National Journals and Conferences. His area of research includes Wireless sensor network and wireless communication.

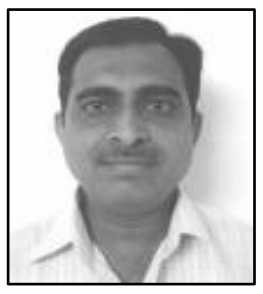

**Dr. V. Kumar** received Ph.D in 1980 and worked as Scientist in CMRI, Dhanbad and Visiting Scientist at CNR, Frascati, Roma, Italy during 1983-84. He has attended number of International Conferences including conferences held in USA, Portugal, Singapore and Italy. At present, he is Associate Professor and Head of Electronic & Instrumentation Department, ISM University, Dhanbad. He has published over 100-research papers; guided 2-Ph.D. and number of M. Tech students in the areas of Opto-electronic materials and Optical Fiber Sensors.

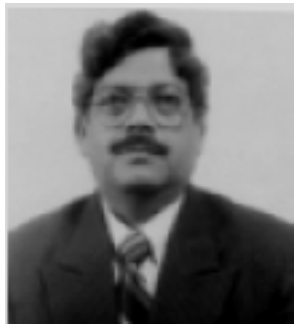

**Dr. B. P. Patil** received his Ph.D. in 2000 and has more than 18 years of teaching and industrial experience. He has published 55 papers in International journals and Conferences also guiding four Ph.D. students. He is currently working as Professor and Head of Electronics Engineering Dept. at Maharashtra Academy of Engineering (MAE), Alandi, and Pune. His area of research includes sensor network and wireless communication.